\documentclass[aps,pre,twocolumn,showpacs,floatfix,superscriptaddress,nofootinbib]{revtex4}
\usepackage{graphicx}
\usepackage{amsmath}

\newcommand {\be} {\begin{equation}}
\newcommand {\ee} {\end{equation}}
\newcommand {\Be}{\begin{eqnarray*}}
\newcommand {\Ee} {\end{eqnarray*}}
\newcommand {\bey} {\begin{eqnarray}}
\newcommand {\eey} {\end{eqnarray}}
\newcommand{\bit}{\begin{itemize}}      
\newcommand{\eit}{\end{itemize}}
\newcommand{\bfl}{\begin{flusleft}}
\newcommand{\efl}{\end{flusleft}}
\newcommand{\bfr}{\begin{flushright}}
\newcommand{\ec}{\end{center}}
\newcommand{\ben}{\begin{enumerate}}    
\newcommand{\een}{\end{enumerate}}

\newcommand{\comment}[1]{}

\begin{document} 

\title{Heterogeneous Mean Field for neural networks with short term plasticity}
\author{Matteo di Volo}
\email{matteo.divolo@fis.unipr.it}
\affiliation{Dipartimento di Fisica e Scienza della Terra,  Universit\`a di Parma, via G.P. Usberti, 7/A - 43124, Parma, Italy}
\affiliation{Centro Interdipartimentale per lo Studio delle Dinamiche Complesse, via Sansone, 1 - 50019 Sesto Fiorentino, Italy}
\affiliation{INFN, Gruppo Collegato di Parma, via G.P. Usberti, 7/A - 43124, Parma, Italy}
\author{Raffaella Burioni}
\email{raffaella.burioni@fis.unipr.it}
\affiliation{Dipartimento di Fisica e Scienza della Terra,  Universit\`a di
Parma, via G.P. Usberti, 7/A - 43124, Parma, Italy}
\affiliation{INFN, Gruppo Collegato di Parma, via G.P. Usberti, 7/A - 43124, Parma, Italy} 
\author{Mario Casartelli}
\email{mario.casartelli@fis.unipr.it}
\affiliation{Dipartimento di Fisica e Scienza della Terra,  Universit\`a di
Parma, via G.P. Usberti, 7/A - 43124, Parma, Italy}
\affiliation{INFN, Gruppo Collegato di Parma, via G.P. Usberti, 7/A - 43124, Parma, Italy} 
\author{Roberto Livi}
\email{livi@fi.infn.it}
\affiliation{Dipartimento di Fisica,  Universit\`a di Firenze, via Sansone, 1 - 50019 Sesto Fiorentino, Italy}
\affiliation{Istituto dei Sistemi Complessi, CNR, via Madonna del Piano 10 - 50019 Sesto Fiorentino, Italy}
\affiliation{INFN Sez. Firenze, via Sansone, 1 -50019 Sesto Fiorentino, Italy}
\affiliation{Centro Interdipartimentale per lo Studio delle Dinamiche
Complesse, via Sansone, 1 - 50019 Sesto Fiorentino, Italy}
\author{Alessandro Vezzani}
\email{alessandro.vezzani@fis.unipr.it}
\affiliation{ S3, CNR Istituto di Nanoscienze, Via Campi, 213A - 41125 Modena, Italy}
\affiliation{Dipartimento di Fisica e Scienza della Terra,  Universit\`a di Parma, via G.P. Usberti, 7/A - 43124, Parma, Italy}
\begin{abstract}
 We report about the main dynamical features of a model of
 leaky-integrate-and fire excitatory neurons  with short term
 plasticity  defined on random massive networks. We investigate the
 dynamics by a Heterogeneous Mean--Field formulation of the model,
 that is able to reproduce dynamical phases characterized by  the
 presence of quasi--synchronous events. This formulation allows one to solve also the inverse problem of
reconstructing the in-degree distribution for different network topologies from the knowledge of  
the global activity field. We study the robustness of this inversion procedure, by
providing  numerical evidence that  the in-degree distribution 
can be recovered also in the presence of noise and disorder in the
external currents. Finally, we discuss the validity of the heterogeneous mean--field approach for 
sparse networks, with a sufficiently large average in--degree.

\end{abstract}
   

\maketitle


\section{Introduction}
Physiological information about neural structure and activity 
was employed from the very beginning to construct effective mathematical models of 
brain functions. Typically, 
neural networks were introduced as assemblies of elementary dynamical units, that interact with 
each other through a graph of connections \cite{boccaletti}. Under the stimulus of experimental investigations,  
these models have been including finer and finer details.
For instance, the combination of complex single--neuron  dynamics, delay and plasticity in
synaptic evolution, endogenous noise and specific  network topologies revealed quite crucial
for reproducing experimental observations, like the spontaneous emergence of synchronized
neural activity,  both {\sl in vitro} (see, e.g.,  \cite{volman}) and  {\sl in vivo},  and the appearance 
of peculiar fluctuations, the so--called ``up--down" states, in
cortical sensory areas \cite{miguel,torres}. 


Since the brain activity is a dynamical process, its statistical
description needs to take into account time as an intrinsic variable. 
Accordingly, non--equilibrium statistical mechanics should be the proper 
conceptual frame, where effective models of collective brain activity should be casted in. 
Moreover, the large number of units and the redundancy of connections suggest that
a mean--field approach can be the right mathematical tool for
understanding the large--scale dynamics of neural network
models. Several analytical and numerical investigations have been devoted to mean field approaches to 
neural dynamics. In particular, stability analysis of asynchronous states in globally coupled networks and collective observables in highly connected sparse network can be deduced in relatively simple neural network models through mean field techniques  \cite{mfbrunel,mfcessac,mfbress,polmf,millman}.

In this paper we provide a detailed account of a mean--field approach, that
has been inspired by the    
``heterogeneous   mean--field" (HMF) formulation,
recently introduced for general interacting networks \cite{vespignani,mendes}.
The overall method is applied here to the simple case 
of random networks of leaky integrate--and--fire (LIF) excitatory neurons in the
presence of synaptic plasticity.  On the other hand,  it can be applied
to a much wider class of neural network models, based on a similar mathematical
structure.

The main advantages of the HMF method are the following:  ({\sl i}) it can identify  the 
relation between the dynamical properties of the global ({\sl synaptic}) activity 
field and the network topology, ({\sl ii})  it allows one to establish under which conditions  partially
synchronized or irregular firing events may appear , ({\sl iii}) it provides a solution to the inverse 
problem of recovering the network structure from the features of the global activity
field.

In Section \ref{sec2}, we describe the network model of excitatory LIF neurons with
short--term plasticity. The  dynamical properties of the model are discussed at
the beginning of  Section \ref{sec3}. In particular, we recall that the random structure of the
network is responsible for  the spontaneous organization of  neurons in two
families of {\sl locked} and {\sl unlocked} ones \cite{DLLPT}. In the rest of this Section  
we summarize how to define a {\sl heterogeneous thermodynamic limit}, that
preserves the effects of the network randomness and allows one to transform
the original dynamical model into its HMF representation
\cite{BCDLV}). The HMF equations provide a relevant computational advantage with respect
to the original system. Actually, they describe the dynamics
of classes of equal--in--degree neurons, rather than that of individual neurons.
In practice, one can take advantage of a suitable sampling, according to its probability
distribution, of the continuous
in--degree parameter present in the HMF formulation.
For instance, by properly "sampling" the HMF model into  300 equations one can 
obtain an effective description of the dynamics engendered by a random
Erd\"os--Renyi network made of ${\mathcal O} (10^4)$ neurons.

In Section  \ref{sec4} we show that the HMF  formulation
allows also for a clear interpretation of the presence of classes of {\sl locked} and {\sl unlocked}
neurons in QSE: they correspond to the presence of a {\sl fixed point} or of an {\sl intermittent-like} map of the 
return time of firing events, respectively. Moreover, we analyze in details the stability properties of the model and we find that any finite sampling of the
HMF dynamics is chaotic, i.e. it is characterized by a positive maximum Lyapunov exponent,
$\lambda_{\mathrm max}$. Its value depends indeed on the finite sampling
of the in--degree parameter. On the other hand,  chaos is found to be relatively weak and, when the number 
of samples, $M$,  is increased,
$\lambda_{\mathrm max}$ vanishes with a power--law decay, $M^{-\gamma}$, with $\gamma \sim 1/2$.
This is  consistent with the mean--field like nature of the HMF equations: in fact, it
can be argued that, in the thermodynamic limit, any chaotic component of the dynamics
should eventually disappear, as it happens for the original LIF model, when a naive
thermodynamic limit is performed \cite{DLLPT}.

In Section \ref{sec5} we analyze the HMF dynamics for networks with different topologies (e.g., Erd\"os--Renyi
and in particular scale free). We find that the dynamical phase characterized by QSE is robust with
respect to the network topology and it can be observed only if the variance of the considered
in--degree distributions is sufficiently small. In fact, quasi-synchronous events are
suppressed for too broad in--degree distributions, thus yielding a transition between 
a fully asynchronous dynamical phase  and a quasi-synchronous one, controlled by the
variance of the in--degree distribution. In all the cases analyzed in this Section, we find
that the global synaptic--activity field characterizes completely
the dynamics in any network topology. 

Accordingly,  the HMF formulation appears as an
effective algorithmic tool 
for solving the following {\sl inverse problem}: given a global
synaptic--activity field, which kind of network topology 
has generated it? In Section \ref{sec6}, after a summary of the
numerical procedure used to solve such an inverse problem, we analyze 
the  robustness of the method in two circumstances: $a)$ when a noise is 
added to the average synaptic--activity field, and $b)$ when there are
noise and disorder in the external currents.

Such robustness studies are particularly relevant in view of applying this strategy to
real data  obtained from experiments.  
Finally, in Section \ref{sec7} we show that a HMF formulation can be 
straightforwardly extended to non--massive networks, i.e. random networks, where the in--degree does not
increase proportionally to the number of neurons. In this case the relevant quantity
in the HMF-like formulation is the average value of the in--degree
distribution, and  the HMF equations are expected to reproduce confidently the dynamics of
non--massive networks, provided this average is sufficiently large.
Conclusions and perspectives are contained in Section \ref{sec8}.

\section{The model}\label{sec2}
We consider a network of $N$ excitatory LIF neurons
interacting via a synaptic current and regulated by short--term plasticity, 
according to a model introduced in \cite{tsodyksnet}. 
The membrane potential $V_j$ of each neuron evolves in time following the
differential equation
\begin{equation}
\label{eq1}
\tau_\mathrm{m} \dot V_j= E_{\mathrm{c}} -V_j + R_\mathrm{in}I_{\mathrm{syn}}(j)\, ,
\end{equation}
where $\tau_\mathrm{m}$ is the membrane time constant, 
$R_{\mathrm{in}}$ is the membrane resistance,
$I_{\mathrm{syn}}(j)$ is the synaptic current received by neuron $j$ from 
all its presynaptic neurons (see below
for its mathematical definition) and 
$E_{\mathrm{c}}$ is the contribution of an  external current 
(properly multiplied by a unit resistance). 
 
Whenever the potential $V_j(t)$ reaches the threshold value $V_{\mathrm{th}}$, it is 
reset to $V_{\mathrm{r}} $, and a spike is sent towards the postsynaptic neurons. 
For the sake of simplicity the spike is assumed to be a $\delta$--like function of time. 
Accordingly, the spike--train $S_j(t)$ produced by neuron $j$, is defined as,
\begin{equation}
\label{eq2}
S_j(t)=\sum_m \delta(t-t_{j}(m)),
\end{equation}
where $t_{j}(m)$ is the time when neuron $j$ fires its $m$-th spike.

The transmission of the  spike--train $S_j(t)$ is mediated by the
synaptic dynamics.
We assume that all efferent synapses of a given neuron follow 
the same evolution (this is justified in so far as no inhibitory 
coupling is supposed to be present).  The state of the $i$-th synapse is
characterized by three variables, $x_i$, $y_i$, and $z_i$, which represent the
fractions of synaptic transmitters in the recovered, active, and inactive state, 
respectively ($x_i+y_i+z_i=1$) \cite{plast1,plast2,tsodyksnet}. 
The evolution equations are
\begin{align}
\label{dynsyn}
& \dot y_{i} = -\frac{y_{i}}{\tau_{\mathrm{in}}} +ux_{i}S_i\\
\label{contz}
& \dot z_{i} = \frac{y_{i}}{\tau_{\mathrm{in}}}  -   \frac{z_{i}}{\tau_{\mathrm{r}}} \ .
\end{align} 
Only the active transmitters react to the incoming spikes: the parameter $u$ 
tunes their effectiveness. Moreover, $\tau_{\mathrm{in}}$ is the characteristic decay time of the
postsynaptic current, while $\tau_{\mathrm{r}}$ is the recovery time from synaptic depression. 
For the sake of simplicity, we assume also that all parameters appearing in the above 
equations are independent of the neuron indices. 
The model equations are finally closed, by representing the synaptic current 
as the sum of all the active transmitters delivered to neuron $j$ 
\begin{equation}
I_{\mathrm{syn}}(j) = \frac{ G}{N}\sum_{i\ne j}  \epsilon_{ij}y_i,
\label{input}
\end{equation}
where $G$ is the strength of the synaptic coupling (that we assume 
independent of both $i$ and $j$), while $\epsilon_{ij}$ is the directed connectivity
matrix whose entries are set equal to 1 or 0 if the presynaptic neuron
$i$ is connected or disconnected with the postsynaptic neuron $j$, respectively. 
Since we suppose  the input resistance
$R_{\mathrm{in}}$  independent of $j$, it can be included into $G$.
In this paper we study the case of excitatory coupling between neurons,
i.e. $G > 0$.  We assume that each neuron
is connected to a macroscopic number, 
${\mathcal O}(N)$, of pre-synaptic neurons: this is the reason why the sum is divided by the 
factor $N$. 
Typical values of the parameters contained in the model have phenomenological
origin \cite{volman,tsodyksnet}. Unless otherwise stated,  we adopt the following set of values: 
$\tau_\mathrm{in} = 6$ ms, 
$\tau_\mathrm{m} = 30$ ms,  $\tau_\mathrm{r} = 798$ ms, 
$V_{\mathrm{r}} = 13.5$ mV, $V_{\mathrm{th}} = 15$ mV, 
$E_{\mathrm{c}} =15.45$ mV,  ${ G} = 45$ mV and $u = 0.5$.
Numerical simulations can be performed much more effectively by introducing
dimensionless quantities,
\begin{align}
& a = \frac{E_c-V_{\mathrm{r}}}{V_{\mathrm{th}}-V_{\mathrm{r}}}\\
& g = \frac{G}{V_{\mathrm{th}}-V_{\mathrm{r}}}\\
&  v=\frac{V-V_\mathrm{r}}{V_{\mathrm{th}}-V_\mathrm{r}},
\end{align}  
and by rescaling time, together with all the other temporal parameters, in units of the membrane time 
constant $\tau_\mathrm{m}$ 
(for simplicity, we leave the notation unchanged after rescaling). The values of the
rescaled parameters are:
$\tau_\mathrm{in} = 0.2$,  $\tau_{\mathrm{r}} = 133\tau_{\mathrm{in}}$, $v_{\mathrm{r}} = 0$, 
$v_{\mathrm{th}} = 1$, $a= 1.3$,  $g = 30$ and $u = 0.5$. 
As to the normalized external current $a$, its value for the first part of our analysis corresponds to the firing regime for neurons.
While the rescaled Eqs. (\ref{dynsyn}) and (\ref{contz}) keep the same form, Eq.~(\ref{eq1}) 
changes to,
\begin{equation}
\label{eq1n}
\dot v_j= a -v_j + \frac{g}{N} \sum_{i \ne j} \epsilon_{ij} y_i \, .
\end{equation}
A major advantage for numerical simulations comes from the possibility of transforming 
the set of differential equations (\ref{dynsyn})--(\ref{input}) and (\ref{eq1n})
into an event--driven map (for details see \cite{DLLPT} and also  \cite{brette,zill}).

\section{Dynamics and heterogeneous mean field limit}\label{sec3}

The dynamics of the fully coupled neural network (i.e., $\epsilon_{ij}=1, \, \forall i,j$),
described by Eq.s (\ref{eq1n}) and  (\ref{eq2})--(\ref{input}), converges to a periodic synchronous state, 
where all neurons fire simultaneously and the period depends on the model parameters \cite{DLLPT}.   
A more interesting dynamical regime appears when some disorder is introduced in the network structure.
For instance, this can be obtained by maintaining each link between neurons with probability
$p$, so that the in-degree of a neuron (i.e.  the number of  presynaptic connections acting on it)
takes the average value $\langle k_i \rangle = p N$, and the
standard deviation of the corresponding in-degree distribution is given by the relation $\sigma_{k} = \sqrt{Np(1-p)}$.
In such an Erd\"os-Renyi random network one typically  
observes quasi--synchronous events (QSE), where a large fraction of neurons  fire in a short
time interval of a few milliseconds, separated by an irregular firing activity lasting over some tens of ms
(e.g., see \cite{DLLPT}). 
This dynamical regime emerges as a collective phenomenon, where neurons separate spontaneously into
two different families: the {\it locked}  and the {\sl unlocked} ones. 
Locked neurons determine the QSE and exhibit a periodic behavior, with a common period but different phases.
Their in--degree $k_i$  ranges over a finite interval below the  average value $\langle k_i \rangle$.
The unlocked ones participate to the irregular firing activity and exhibit a sort of intermittent evolution \cite{DLLPT}.
Their in-degree is either very small or higher than  $\langle k_i \rangle$.


As the dynamics is very sensitive to the different values of of $k_i $, in a recent publication \cite{BCDLV} we have  shown that one can design a
{\it heterogeneous mean-field} (HMF) approach by  a suitable
thermodynamic limit preserving, for increasing values of $N$,  the main features associated with topological disorder.  
The basic step of this approach is the
introduction of a  probability distribution,  $P(\tilde k)$, for the
normalized in-degree variable ${\tilde k} = k/N$, where the average $\langle \tilde k \rangle$
and the variance $\sigma_{\tilde k}^2 = \langle {\tilde k}^2 \rangle - \langle \tilde k \rangle^2$ 
are fixed independently of $N$.  
A realization of the random network containing $N$ nodes (neurons) 
is obtained by extracting for each neuron $i$ ($i =1, \cdots , N$) a value $\tilde k_i$ from $P(\tilde k)$, and 
by connecting the neuron $i$ with $\tilde k_i  N$ randomly chosen neurons (i.e., $\epsilon_{i,j} = 1$,  $j(i)= 1, \cdots , \tilde k_i N $). 
For instance, one can consider a suitably normalized Gaussian--like  distribution 
defined on the compact support, $\tilde k \in (0,1]$,
centered around  $\langle \tilde k \rangle$ with a sufficiently small value of the standard deviation  
$\sigma_{\tilde k}$, so that the tails of the distribution vanish at the boundaries of the support.
\begin{figure}[htbp]
\centering
\includegraphics[scale=0.35]{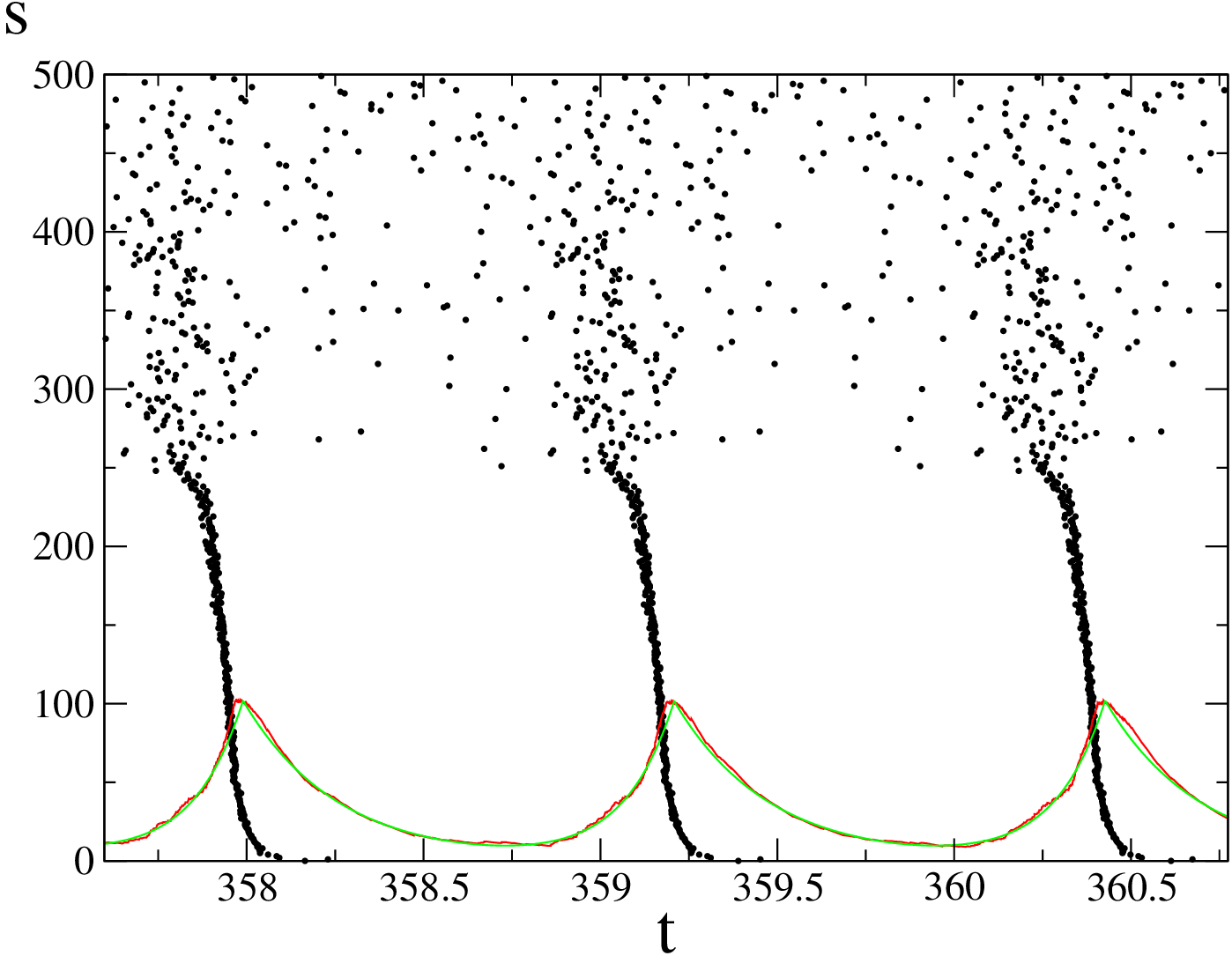}
\caption{
Raster plot of a randomly diluted network containing 500 
neurons, ordered along the
vertical axis according to their in--degree. The distribution $P(\tilde k)$ is a Gaussian with  $\langle \tilde k \rangle =0.7$, standard deviation $\sigma_{\tilde k} =0.077$. A black dot in the raster plot indicates that
neuron $s$ has fired at time $t$. The red line is the global field $Y(t)$ and the green curve is its analytic fit by the function $Y_f(t)=
Ae^{-\frac{t}{\tau_1}}+B(e^{\frac{t}{\tau_2}}-1)$, that repeats over each 
period of $Y(t)$; the parameter values are $A=2 \cdot 10^{-2}$, $B=3.56\cdot 10^{-6}$, $\tau_1=0.268$ and $\tau_2=0.141$. Notice that the amplitude of both $Y(t)$ and $Y_f(t)$ has been suitably rescaled to be appreciated on the same scale of the Raster plot.}
\label{rp1}
\end{figure}

In  Fig.\ref{rp1} we show the raster plot for a network of $N=500$ neurons and a Gaussian distribution $P(\tilde k)$ 
with  $\langle \tilde k \rangle=0.7$ and  $\sigma_{\tilde k}=0.077$. One can observe
a quasi-synchronous dynamics characterized by the presence of locked
and unlocked neurons, and such a distinctive dynamical feature is preserved
in the thermodynamic limit \cite{BCDLV}. For example the time average of the inter--spike time interval between firing events of each neuron, (in formulae $ISI_m=t_m-t_{m-1}$, where the integer $m$ labels the $m$-th firing event)  as a function of the connectivity $\tilde k$ is, apart from fluctuations, the same for each network size $N$. This confirms that the main features of the dynamics are maintained for increasing values of $N$.


 The main advantage of this approach is that one can 
explicitly perform the limit $N\to \infty$ on the set of equations
(\ref{eq1n}) and (\ref{eq2})--(\ref{input}), thus
obtaining the corresponding HMF equations:

\begin{align}
\label{vk}
&\dot v_{\tilde k}(t)= a -v_{\tilde k}(t) + g\tilde kY(t)\\
\label{sk}
&S_{\tilde k}(t) = \sum_m \delta(t-t_{\tilde k}(m)) \\
\label{yk}
& \dot y_{\tilde k}(t) = -\frac{y_{\tilde k}(t)}{\tau_{\mathrm{in}}} +u(1-y_{\tilde k}(t)-z_{\tilde k}(t))S_{\tilde k}(t)\\
\label{zk}
& \dot z_{\tilde k}(t) = \frac{y_{\tilde k}(t)}{\tau_{\mathrm{in}}}  -   \frac{z_{\tilde k}(t)}{\tau_{\mathrm{r}}}\\
\label{meanfield}
&Y(t)=\int_{0}^{1}P(\tilde k) y_{\tilde k}(t)d\tilde k .
\end{align}
The dynamical variables depend now on the continuous  in--degree index $\tilde k$, and
this set of equations represents the dynamics of equivalence classes of neurons. In fact, in this HMF formulation,
neurons with the same $\tilde k$  follow the same evolution
\cite{vespignani, mendes}. In practice,  Eq.s   (\ref{vk})--(\ref{meanfield}) can be integrated numerically by  sampling  the 
probability distribution $P(\tilde k)$: one can subdivide  the support $(0,1]$ of $\tilde k$ 
by $M$ values  $\tilde k_i \,\,\, (i=1,\cdots , M)$, in such a way that $\int_{\tilde k_i}^{\tilde k_{i+1}}P(\tilde k)d\tilde k$ is 
constant (importance sampling).  Notice that the integration of the discretized  HMF equations is much less
time consuming than the simulations performed on a random network.
For instance, numerical tests indicate that 
the dynamics of a network with $N=10^4$ neurons can be confidently reproduced by an importance sampling with $M= 300$.

\begin{figure}[htbp]
\centering
\includegraphics[scale=0.35]{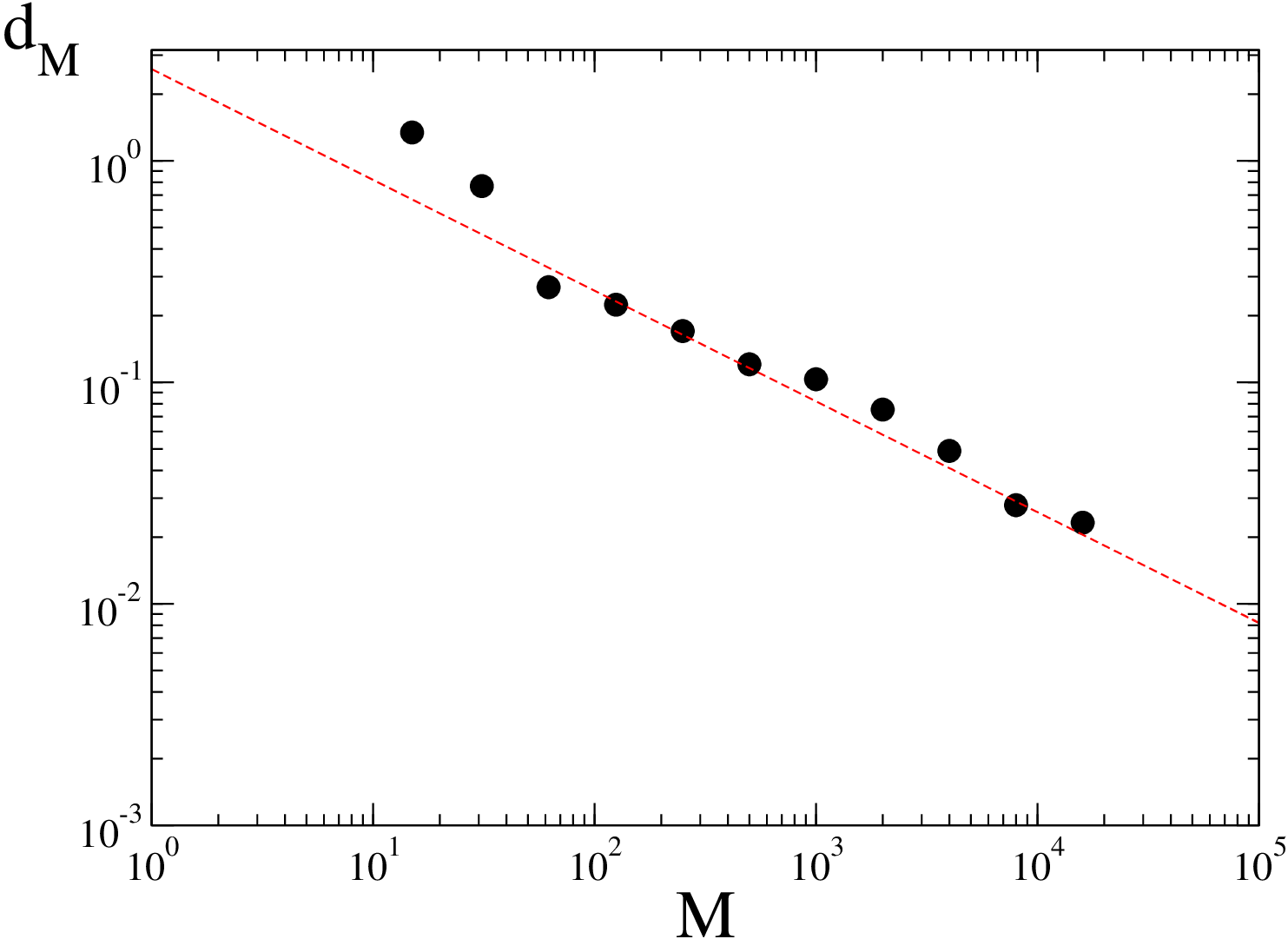}
\caption{(Color online) The effect of sampling the probability distribution
$P(\tilde k)$ with $M$ classes of neurons in the HMF dynamics. Finite size
effects are controlled by plotting the distance
between the activity fields obtained for two sampling values $M$ and $M/2$, $d_M=d(Y_M(t),Y_{M/2}(t))$ (defined in the text), vs. $M$. The red dashed line is 
the power law $1/\sqrt M$. Data is obtained for a Gaussian distribution $P(\tilde k)$, with  $\langle \tilde k \rangle=0.7$ and  $\sigma_{\tilde k}=0.077$.}
\label{scarto}
\end{figure}

The effect of the discretization of ${\tilde k}$ on the HMF dynamics can be analyzed
by  considering the distance $d(Y_{M_1}(t),Y_{M_2}(t))$ between the global
activity fields $Y_{M_1}(t)$ and $Y_{M_2}(t)$ (see Eq.(\ref{meanfield})) obtained for two different values $M_1$ and $M_2$ of the sampling,
 i.e.:
\begin{equation}
d(Y_{M_1}(t),Y_{M_2}(t))=\Bigg(\frac{1}{T}\sum_{i =1}^T\frac{(Y_{M_1}(t_i)-Y_{M_2}(t_i))^2}{Y_{M_1}(t_i)^2}\Bigg)^{\frac{1}{2}}.
\end{equation}
In general $Y(t)$ exhibits a quasi periodic behavior and  $d(Y_{M_1}(t),Y_{M_2}(t))$ is evaluated over a  time interval equal to its period $T$.
In order to avoid an overestimation of  $d(Y_{M_1}(t),Y_{M_2}(t))$
due to different initial conditions, the field  $Y_2(t)$ is suitably translated in time in order to make its
first maximum coincide with the first maximum of  $Y_1(t)$ in the time interval $[1,T]$. 
In Fig. \ref{scarto}  we plot $d_M=d(Y_M,Y_{M/2})$ as a function of $M$. We  find that $d_M\sim 1/\sqrt M$,
thus confirming that the finite size simulation of the HMF dynamics is consistent with the HMF model ($M\to \infty$).

As a final remark, notice that the presence of  short--term synaptic plasticity 
plays a fundamental role in determining the partially synchronized regime.
In fact, numerical simulations show that the discretized  HMF dynamics without plasticity, 
i.e. $Y(t) = \int_{0}^{1}P(\tilde k) S_{\tilde k}(t)d\tilde k$, 
converges to a synchronous periodic dynamics for any value of $M$  \cite{DL} .

\section{Stability analysis of the HMF dynamics}
\label{sec4}

In the HMF equations (\ref{vk})--(\ref{meanfield}) the dynamics of each neuron is
determined by its in--degree $\tilde k$ and by the global synaptic activity field $Y(t)$.
For the stability analysis of these equations, we follow  a procedure 
introduced in \cite{tso_locked} and employed also in \cite{BCDLV}.
For  sufficiently large $M$ the  discretized  HMF dynamics allows one to 
obtain a precise fit of the periodic function $Y(t)$ and to estimate its period $T$. 
As an instance of its  periodic behavior,  
in  Fig.\ref{rp1} we report also $Y(t)$ (red line)  and its fit (green line and the
formula in the caption). The fitted field is exactly periodic and is a good approximation of the global field that
 one expects to observe in the mean field model corresponding to an infinite discretization $M$. As a result, the analysis performed using this periodic field are relative to the dynamics of the HMF model, i.e. in the limit  $M\to \infty$.
Using this fit, one can represent the dynamics of  each class $\tilde k$ of neurons 
by the discrete--time map 
\begin{equation}
\label{mappa}
\tau_{\tilde k}(n+1)=R_{\tilde k}[ \tau_{\tilde k}(n)],
\end{equation}
where $\tau_{\tilde k}(n) = | t_{\tilde k}(n) - nT |$ is  the modulus of the time difference 
between the  $n$-th  spike  of neuron $\tilde k$ and $nT$, i.e. the $n$-th QSE, that 
is conventionally identified by  the corresponding maximum of  $Y(t)$ (see Fig. \ref{rp1}). 

\begin{figure}[htbp]
\centering
\includegraphics[scale=0.35]{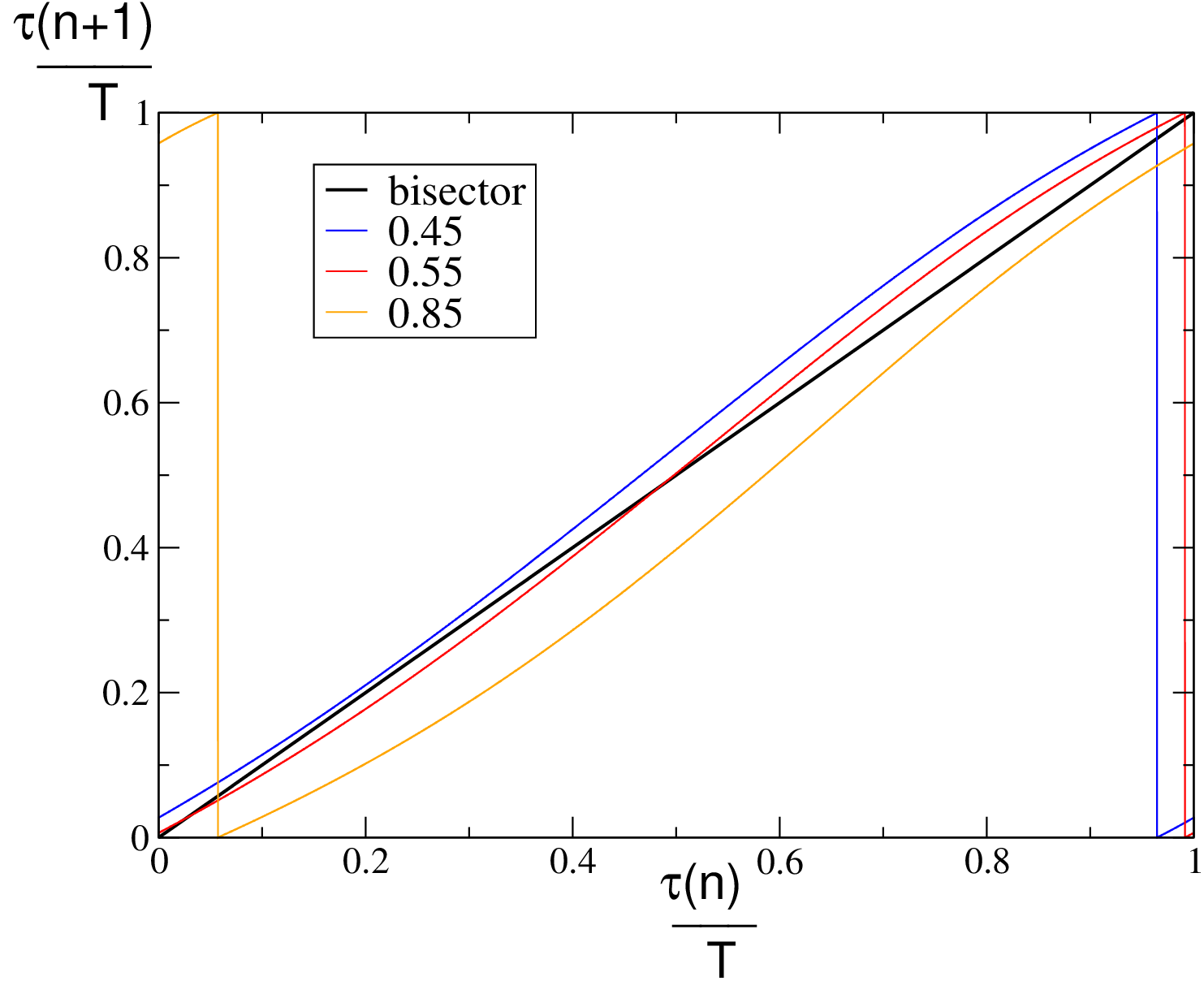}
\caption{The return map $R_{\tilde k}$ of the rescaled variable 
$\tau_{\tilde k}/T$ (see Eq.(\ref{mappa})) for different values of 
$\tilde k$, corresponding 
to lines of different colors (see the legend in the inset: the black line 
is the bisector of the square).
}
\label{maps}
\end{figure}

In Fig. \ref{maps} we show $R_{\tilde k} $ for different values of $\tilde k$. The map
of each class of {\sl locked} neurons has a stable fixed point, whose value decreases with 
$\tilde k$.  As a consequence, different classes of {\sl locked} neurons share the
same periodic behavior, but  exhibit different phase shifts
with respect to the maximum of $Y(t)$. This analysis describes in a clear 
mathematical language what is observed in simulations (see Fig \ref{rp1}):
equally periodic classes of locked neurons determine
the QSE by firing sequentially, over a very short time interval, that depends on
their relative phase shift.
In general, the values of $\tilde k$ identifying the family of locked neurons
belong to a subinterval $(\tilde k_1, \tilde k_2)$ of $(0,1]$: the values of 
$\tilde k_1$  and $\tilde k_2$ mainly depend on $P(\tilde k)$ and on its 
standard deviation $\sigma_{\tilde k}$ (more details are reported in \cite{BCDLV}).
For what concerns {\sl unlocked} neurons,  $R_{\tilde k} $ exhibits the features of an intermittent-like dynamics. In fact, unlocked neurons with $\tilde k$
close to  $\tilde k_1$  and $\tilde k_2$ spend a long time in an almost periodic
firing activity, contributing to a QSE, then they depart from it,  firing  irregularly
before possibly coming back again close to a QSE. The duration of the irregular
firing activity of unlocked neurons typically increases for values of $\tilde k$ far from
the interval $ (\tilde k_1, \tilde k_2 )$.

Using the deterministic map (\ref{mappa}), one can tackle in full
rigor the stability problem of the HMF model. The existence of stable
fixed points for the locked neurons implies that they yield a negative
Lyapunov exponent associated with their periodic evolution.

\begin{figure}[htbp]
\centering
\includegraphics[scale=0.35]{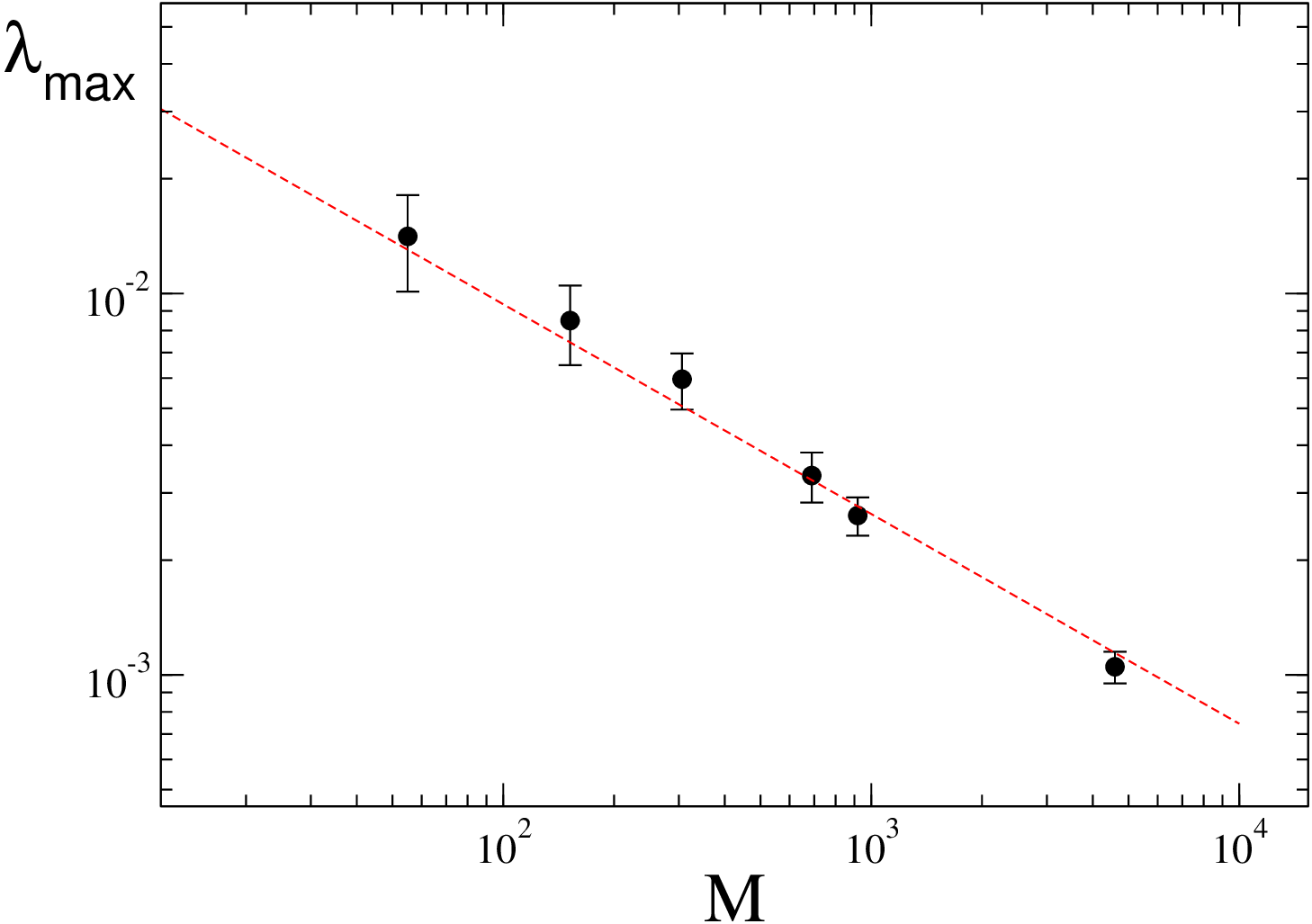}
\caption{(Color online) The maximum Lyapunov exponent $\lambda_{\mathrm{max}}$ 
as a function of the sampling parameter $M$: $\lambda_{\mathrm{max}}$ has
been averaged also over ten different realizations of the network (the error bars refer to the maximum deviation from the average). The dashed red line is the powerlaw  $M^{-\gamma}$, with $\gamma=0.55$. }
\label{lyup_max}
\end{figure}

As for the unlocked neurons,  their  Lyapunov
exponent, $\lambda_{\tilde k}$,  can be calculated numerically by 
 the time-averaged  expansion rate 
of nearby orbits  of  map (\ref{mappa}):
\begin{equation}
\label{lyn}
\lambda_{\tilde k}(n)=  \frac{1}{n} \sum_{j=1}^n \mathrm{log}\Bigg[\frac{|\delta(j)|}{|\delta(0)|}\Bigg],
\end{equation}
where $\delta(0)$ is the initial distance between nearby orbits and
$\delta(j)$ is their distance at the $j$--th iterate,  so that
\begin{equation}
\label{lyap}
\lambda_{\tilde k}= \lim_{n \to \infty} \lambda_{\tilde k}(n)
\end{equation}
if this limit exists. The Lyapunov exponents for the unlocked component vanish as
$\lambda_{\tilde k} (n) \sim 1/n$.
 According to these results, one expects that the maximum Lyapunov exponent
 $\lambda_{\mathrm{max}}(M)$  goes to zero in the limit $M \to \infty$. 
In fact, at each finite $M$,  $\lambda_{\mathrm{max}}$ can be evaluated by using the standard algorithm by Benettin et al.
\cite{BGGS}. 
In Fig.\ref{lyup_max} we plot $\lambda_{\mathrm{max}}$  as a function of the discretization parameter $M$. 
Thus, $\lambda_{\mathrm{max}}(M)$ is positive,
behaving approximately  as  $M^{-\gamma}$, with $\gamma \sim 1/2$
(actually,  we find $\gamma = 0.55$).

The  scenario in any discretized version of the HMF dynamics is the following: 
{\sl (i)}  all {\sl unlocked neurons} exhibit positive Lyapunov exponents, i.e. they represent
the chaotic component of the dynamics; {\sl (ii)}  $\lambda_{\mathrm{max}}$ is typically 
quite small, and its value depends on the discretization parameter
$M$ and on $P(\tilde k)$; {\sl (iii)} in the limit $M\to \infty$  $ \lambda_{\mathrm{max}}$ and
all $\lambda_{\tilde k}$'s of unlocked neurons vanish, thus converging to a quasi periodic
dynamics, while the {\sl locked neurons} persist in their periodic behavior.

The same scenario is observed in the dynamics of random networks built with the
HMF strategy, where the variance of the distribution $P(\tilde k)$ is kept independent of the 
system size $N$, so that  the fraction of locked neurons is constant.

For the LIF dynamics in an E\"ordos--Renyi random network with $N$ neurons, it was found that 
$ \lambda_{\mathrm{max}}(N) \approx N^{- 0.27}$  in the limit $N\to\infty$ \cite{DLLPT}. According to the argument proposed in \cite{DLLPT},  the value of the power-law exponent is associated to the
 scaling of the  number of unlocked neurons, $N_u$  with the system size $N$, namely $N_u \sim N^{0.9}$.
The same argument applied to HMF dynamics indicates that the exponent 
$\gamma \sim 1/2$, ruling the vanishing of $\lambda_{\mathrm{max}} (M)$ in the limit $M\to\infty$,
stems from the fact that the HMF dynamics keeps the fraction of unlocked neurons constant.

\vskip 30pt

When the distribution $P(\tilde k)$ is sufficiently broad, the system becomes asynchronous and locked neurons disappear. The global field  $Y(t)$ exhibits  fluctuations due to finite size effects and in the thermodynamic limit 
it tends to a constant value  $Y^*$. From Eq.s (\ref{vk})--(\ref{zk}), one obtains that in this regime each neuron
with in--degree $\tilde k$ fires  periodically with a period 
 $$
 T_{\tilde k}=\mathrm{ln}\Bigg[\frac{b+g\tilde k Y^*}{b+g\tilde k Y^* -1}      \Bigg]~,
 $$
while its  phase depends  on the initial conditions. In this case all the Lyapunov exponents 
$\lambda_{\tilde k}$ are negative.

\section{Topology and collective behavior}
\label{sec5}
For a  given in--degree probability distribution
$P(\tilde k)$, the fraction of locked neurons (i.e.,  $f_{\mathrm{l}}=\int_{\tilde k_1 }^{\tilde k_2} P(\tilde k)d\tilde k$)  decreases 
by increasing $\sigma_{\tilde k}$ \cite{BCDLV}. In particular, there is a critical value  $\sigma^*$ at which $f_{\mathrm{l}}$ vanishes.
This signals a very interesting dynamical transition between the quasi-synchronous phase ($\sigma_{\tilde k} < \sigma^*$)
to a multi-periodic phase ($\sigma_{\tilde k} > \sigma^*$), where all neurons are periodic with different periods. Here we focus on the different  collective dynamics that may emerge for choices of  $P(\tilde k)$
other than the Gaussian case, discussed in the previous section.
 
First, we consider a power--law distribution
\begin{equation}
P(\tilde k)=A\tilde k^{-\alpha},
\label{eqplaw}
\end{equation}
where  the constant $A$ is given  by the normalization condition $\int_{\tilde k_m}^{1}P(\tilde k)d\tilde k=1$. 
The lower bound $\tilde k_m$ is introduced  in order to maintain $A$ finite. 
For simplicity, we fix  the parameter $\tilde k_m$ and analyze the dynamics by varying $\alpha$. 
Notice that the standard deviation  $\sigma_{\tilde k}$ of  distribution (\ref{eqplaw})  decreases  for increasing
values of  $\alpha$.  The dynamics for relatively high $\alpha$  is very similar to the quasi--synchronous regime 
observed for $\sigma_{\tilde k} < \sigma^*$  in the Gaussian case (see Fig. \ref{rp1}). 
By decreasing $\alpha$ one can observe again a transition to  the  asynchronous phase observed 
for  $\sigma_{\tilde k} > \sigma^*$ in the Gaussian case.  Accordingly, also for the power--law distribution (\ref{eqplaw})
a phase with locked neurons may set in only when there is  a sufficiently large group of neurons sharing  close values of $\tilde k$.
In fact, the group of locked neurons is concentrated at values of $\tilde k$ quite close to the lower bound  
 $\tilde k_m$,  while in the Gaussian case they concentrate at values smaller than  $\langle \tilde k\rangle $.
 
Another distribution, generating an  interesting dynamical phase, is
\begin{equation}
P(\tilde k)=B \mathrm{exp}\Bigg(-\frac{(\tilde k-p_1)^2}{2\sigma_s^2}\Bigg) +    B\mathrm{exp}\Bigg(-\frac{(\tilde k-p_2)^2}{2\sigma_s^2}\Bigg)   ,
\label{dgauss}
\end{equation}
i.e. the sum of two Gaussians  peaked around different values, $p_1$ and $p_2$, of $\tilde k$, 
with the same variance  $\sigma_s^2$.  $B$ is the  normalization constant such that $\int_{0}^{1}P(\tilde k)=1$.
We fix $p_1=0.5$ and vary both the variance,
$\sigma_s$, and the distance between the peaks, $\Delta= |p_2-p_1|$.
 
If  $\sigma_s$ is very large ($\sigma \gtrsim 0.1$), the situation is the same observed for a single Gaussian with large variance, yielding  
a multi--periodic asynchronous dynamical phase.

For intermediate values of  $\sigma_s$ i.e. $0.05\lesssim\sigma
\lesssim 0.1$, the dynamics of the network can exhibit 
a quasi--synchronous phase or  a multi--periodic asynchronous phase, depending on the value of $\Delta$.
In fact, one can easily realize that this parameter tunes the standard deviation of the overall distribution: small separations
amount to broad distributions. 
 
\begin{figure}[htbp]
\centering
\includegraphics[scale=0.35]{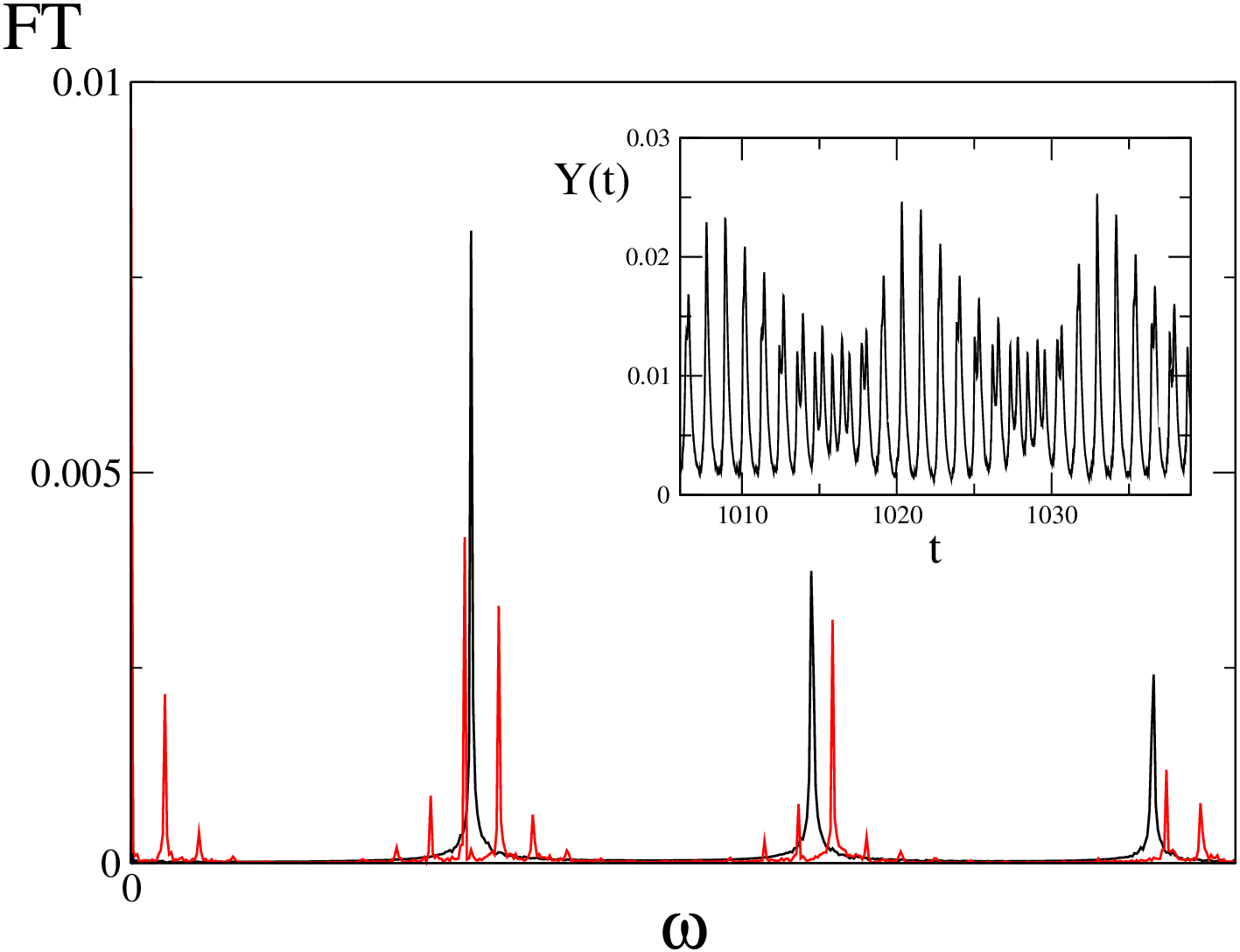}
\caption{The frequency spectra of the global activity field $Y(t)$ for
different in--degree probability distributions. The 
black spectrum has been obtained for the HMF dynamics with $M=350$, generated 
by the power law probability distribution $P(\tilde k)\sim\tilde k^{-4.9}$
(see Eq.(\ref{eqplaw})),
with $\tilde k_m=0.1$: in this case there is a unique family of locked neurons
generating a periodic global activity field $Y(t)$.
The red spectrum has been obtained for a random network of $N = 300$ neurons
generated by the double Gaussian distribution
(see Eq.(\ref{dgauss})) described in Fig.s 6 and 7: in this case two families
 of locked neurons are present while, as reported in the inset, $Y(t)$ exhibits a quasi--periodic 
evolution.}
\label{fourier}
\end{figure}

Finally, when $\sigma_s\lesssim 0.05$, a new dynamical phase appears.  
\begin{figure}[htbp]
\centering
\includegraphics[scale=0.35]{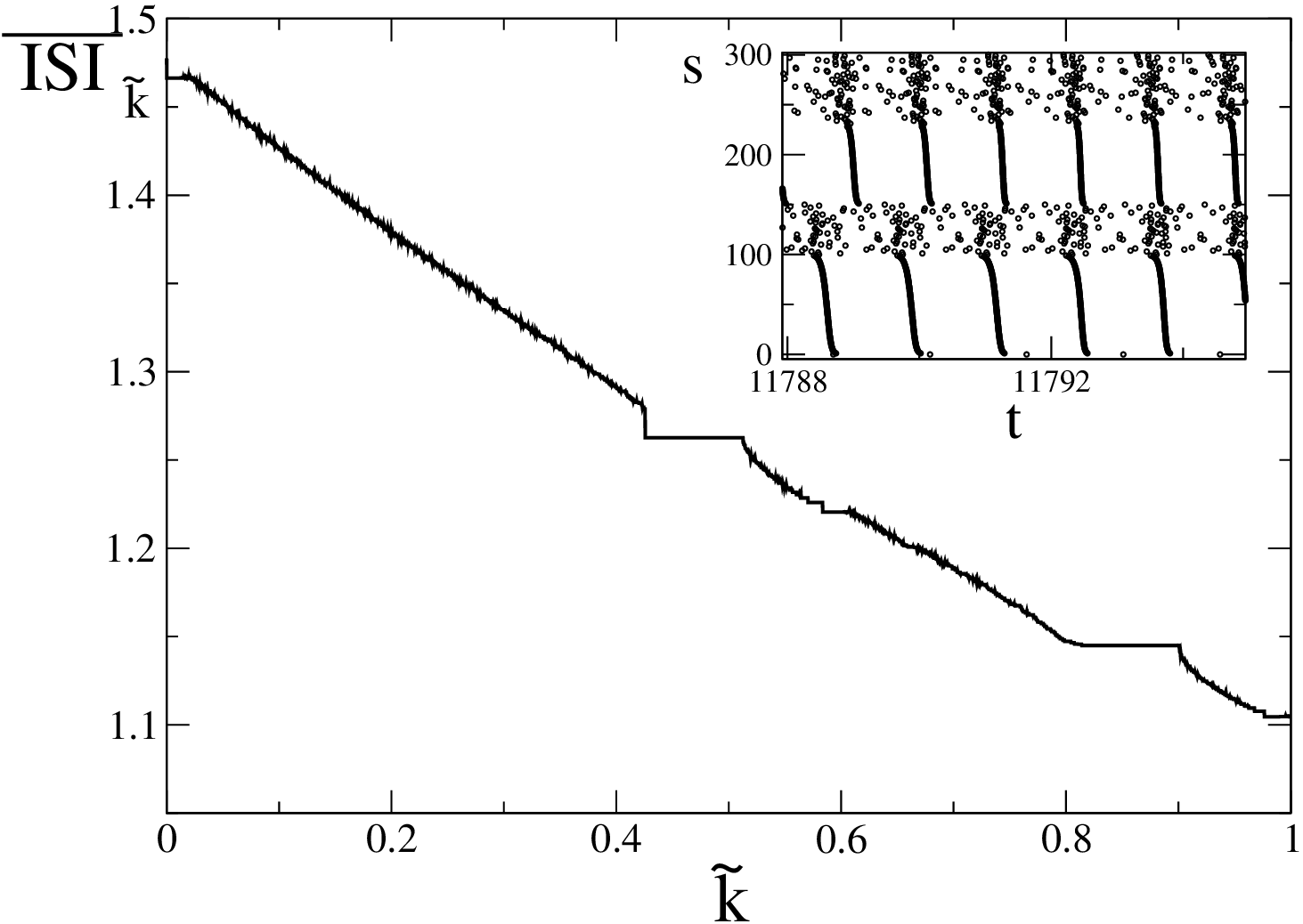}
\caption{ The time average of the inter--spike interval 
$\overline{ ISI_{\tilde k}}$ vs. $\tilde k$ for the probability distribution
$P(\tilde k)$ defined in Eq.(\ref{dgauss}), with  $\Delta= |p_2- p_1|= 0.4$, and $\sigma_{\mathrm{s}}=0.03$. We have obtained the global field $Y(t)$ simulating the HMF dynamics with a discretization with $M=300$ classes of neurons. We have then used $Y(t)$ to calculate the $ISI$ of neurons evolving Eq. (\ref{vk}). In the inset we show the raster plot of the dynamics: as
in Fig.1, neurons are ordered along the vertical axis according to their 
in--degree.
}
\label{rpdue}
\end{figure} 
For small values of $\Delta$ (e.g. $\Delta \approx 0.1$) , we observe the usual QSE scenario with one family of locked neurons
(data not shown). However, when $\Delta$ is sufficiently large
(e.g. $\Delta \approx  0.4$), each peak of the distribution generates its own group of locked neurons.  More precisely, neurons separate into three different sets: 
two locked groups, that evolve with different periods, $T_1$ and  $T_2$, and the unlocked group. In Fig.\ref{rpdue} we show the dependence of $\overline{ ISI_{\tilde k} }$ on $\tilde k$  and the raster plot of the dynamics (see the inset)
for $\sigma_s=0.03$ . Notice that  the plateaus of locked neurons extend over values
of $\tilde k$ on the left of  $p_1$ and $p_2$.   
In the inset of Fig. \ref{fourier} we plot  the global activity field $Y(t)$:  the peaks signal the quasi-synchronous firing events  of the two groups of locked neurons. One can
also observe that very long oscillations are present over  a time scale much larger than $T_1$ and $T_2$. They are the effect of the {\sl firing synchrony} of the of two locked families.  In fact, the two frequencies $\omega_1=2\pi/T_1$ and $\omega_2=2\pi/T_2$ are in general not commensurate, and the resulting global field is a quasi--periodic function. 
This can be better appreciated by looking at Fig.\ref{fourier},  where we report the frequency spectrum of the signal $Y(t)$ (red curve). We observe peaks at frequencies $\omega=n\omega_1+m\omega_2$, for integer values of  $n$ and $m$. For comparison, we report also the spectrum of a periodic $Y(t)$, generated by the HMF with power law probability distribution 
(\ref{eqplaw}), with  $\alpha = 4.9$ (black curve):  in this case the peaks are  located at frequencies multiples of the frequency of the locked group of neurons.

On the basis of this analysis, we can conclude that slow oscillations of the global activity field $Y(t)$ may signal the presence 
of more than one group of topologically homogeneous (i.e. locked) neurons. Moreover, we have also learnt
 that one can generate a large variety of global synaptic activity fields by selecting suitable
in-degree distributions $P(\tilde k)$, thus unveiling unexpected perspectives for exploiting a sort of  {\sl  topological engineering} of the 
neural signals. For instance, one could investigate which kind of $P(\tilde k)$ could give rise to an almost resonant dynamics,
where  $\omega_2$ is close to a multiple of  $\omega_1$. 

\section{HMF and the Inverse problem in presence of noise}
\label{sec6}
The HMF  formulation allows  one to define and solve the following
global inverse problem: how to recover  the 
in--degree distribution $P(\tilde k)$ from the knowledge of  the global synaptic activity field  $Y(t)$ \cite{BCDLV}.  

Here we just sketch the basic steps of the procedure.
Given $Y(t)$, each class of neurons of in-degree $\tilde k$ evolves according to the HMF equations:
\begin{align}
\label{vktil}
&\dot {\mathcal{ V}}_{ \tilde k}(t)= a -\mathcal{ V}_{ \tilde k}(t) + g \tilde kY(t)\\
\label{yktil}
& \dot {\mathcal{ Y}}_{ \tilde k}(t) = -\frac{\mathcal{Y}_{\tilde k}(t)}{\tau_{\mathsf{in}}} +u(1-\mathcal{ Y}_{ \tilde{k}}(t)-\mathcal{ Z}_{ \tilde{k}}(t))\tilde S_{ \tilde{k}}(t)\\
\label{zktil}
& \dot {\mathcal{ Z}}_{ \tilde k}(t) = \frac{\mathcal{ Y}_{ \tilde k}(t)}{\tau_{\mathsf{in}}}  -   \frac{\mathcal{ Z}_{ \tilde{k}}(t)}{\tau_{\mathsf{r}}} \,\,\, .
\end{align}
The different fonts used here, with respect to 
Eq.s   (\ref{vk})--(\ref{meanfield}), point out that
in this framework the choice of the initial conditions is arbitrary and the dynamical variables $\mathcal{V}(t)$, $\mathcal{Y}(t)$, $\mathcal{Z}(t)$ 
in general may take different values  from those assumed by  $v(t)$,
$y(t)$, $z(t)$, i.e. the variables   generating $Y(t)$ in (\ref{vk})--(\ref{meanfield}).  
However,  one can exploit the self consistent relation for the global field $Y(t)$:
\begin{equation}\label{global}
Y(t)=\int_0^1 P({\tilde k}) \mathcal{Y}_{\tilde k}(t)d{\tilde k} \,\,\, .
\end{equation}
If $Y(t)$ and $\mathcal{Y}_{\tilde k}(t)$ are known, this is a Fredholm
equation of the first kind for the unknown $P(\tilde k)$ \cite{kress}.
If $Y(t)$ is a periodic signal,  Eq. (\ref{global}) can be easily solved by a functional Montecarlo minimization
procedure, yielding a faithful reconstruction of  $P(\tilde k)$
\cite{BCDLV}.
This method applies successfully also  when $Y(t)$ is a quasi-periodic signal,  like the
one generated by in--degree distribution (\ref{dgauss}).

In this section we want to study the robustness of the HMF equations and 
of the corresponding inverse problem procedure in the presence of noise. 
This is quite an important test for the reliability of the overall HMF approach. 
In fact, a real neural structure is always affected by some level of noise, that,
for instance, may emerge in the form of fluctuations of ionic or synaptic currents.  
Moreover, it has been observed that noise is crucial for reproducing  dynamical phases,
that exhibit  some peculiar synchronization patterns observed in  {\it in vitro}  experiments \cite{volman,DL}.

For the sake of simplicity, here we introduce noise by turning the
external current  $a$, in Eq. (\ref{vk}), from a  constant to a time  and neuron dependent stochastic processes
$a_{\tilde k}(t)$. Precisely, the $a_{\tilde k}(t)$ are assumed to be 
i.i.d.  stochastic variables, that evolve in time as a random walk with boundaries, 
$a_{\mathrm{min}}$ and $a_{\mathrm{max}}$ (the same rule adopted in \cite{DL}).
Accordingly, the average value, $\bar a$ of  $a_{\tilde k}(t)$ is given by the expression $\bar
a=(a_{\mathrm{min}}+a_{\mathrm{max}})/2$, while the amplitude of fluctuations is
$\delta = a_{\mathrm{max}}-a_{\mathrm{min}}$.
At each step of the walk, the values of $a_{\tilde k}(t)$ are independently updated by adding or subtracting, with equal
probability, a fixed increment $\Delta a$. Whenever the value of $a_{\tilde k}(t)$ crosses one of the boundaries, it is reset to the boundary value.

Since the dynamics has lost its deterministic character,  its numerical integration cannot exploit an event driven algorithm, and 
one has to integrate Eq.s (\ref{vk}) --(\ref{zk}) by a scheme based on
explicit time discretization. The results reported hereafter 
refer to an integration time step  $\Delta t=9\cdot 10^{-4}$, that guarantees an effective sampling of the dynamics over the whole
range of parameter values that we have explored. We have assumed that $\Delta t$  is also the time step of the stochastic evolution of $a_{\tilde k}(t)$.

Here we consider the case of uncorrelated noise, that can be obtained by a suitable choice of $\Delta a$
\cite{DL}. In our simulations $\Delta a = 10^{-2}$, that yields 
a value $\mathcal{O}(10^{-2})$ of the  correlation time of the random walk with boundaries. This value, 
much smaller than the value  $\mathcal{O}(1)$ typical of the  ISI of neurons, makes the stochastic evolution
of the external currents, $a_{\tilde k}(t)$, an effectively  uncorrelated process with respect to the typical time scales of the neural
dynamics.
\begin{figure}[htbp]
\centering
\includegraphics[scale=0.35]{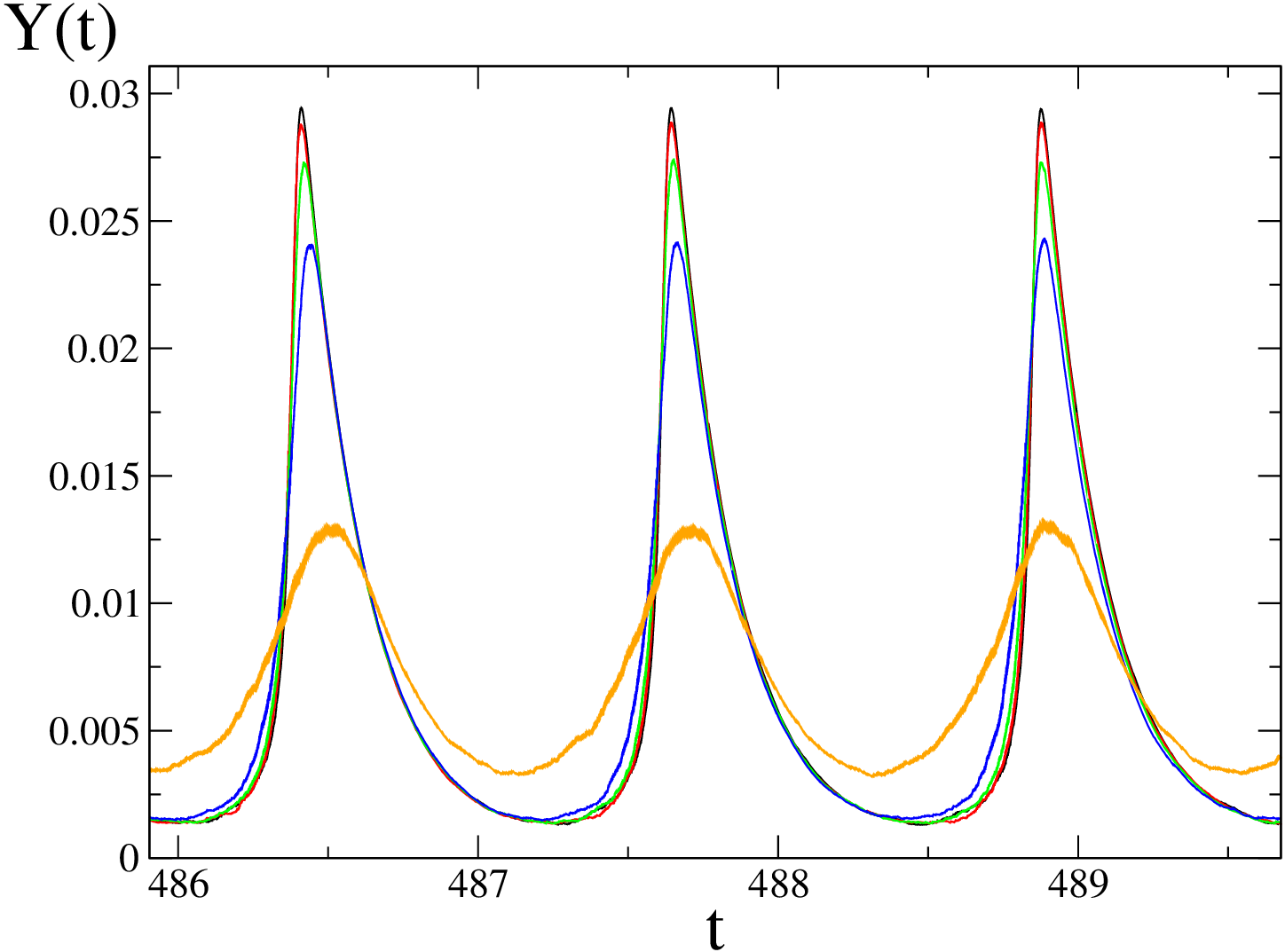}
\caption{ The global activity field $Y(t)$ of the HMF dynamics, sampled by
$M=4525$ classes of neurons, for a gaussian
probability distribution $P(\tilde k)$, with $\langle \tilde k\rangle=0.7$ 
and $\sigma_{\tilde k}=0.0455$. Lines of different colors correspond to
different values of the noise amplitude, $\delta$, added to the external 
currents $a_{\tilde k}(t)$: $\delta = 0$ (black line), $\delta = 0.1$ 
(red line), $\delta = 0.15$ (green line), $\delta = 0.2$ (blue line) and
$\delta = 0.3$ (orange line).}
\label{camponoise}
\end{figure} 
In Fig. \ref{camponoise} we show  $Y(t)$, produced by the discretized HMF dynamics with 
$M=4525$ and for a Gaussian distribution $P(\tilde k)$, with  $\langle \tilde k \rangle=0.7$ and $\sigma_{\tilde k}=0.0455$. 
Curves of different colors correspond to different values of $\delta$. 
 We have found that up to  $\delta\simeq 0.1$, i.e. also for non negligible noise
amplitudes ($\bar a =1$),  the HMF dynamics is practically unaffected by noise.
By further increasing $\delta$, the amplitude of $Y(t)$   decreases, as a result
of the desynchronization of the network induced by large amplitude noise.

\begin{figure}[htbp]
\centering
\includegraphics[scale=0.35]{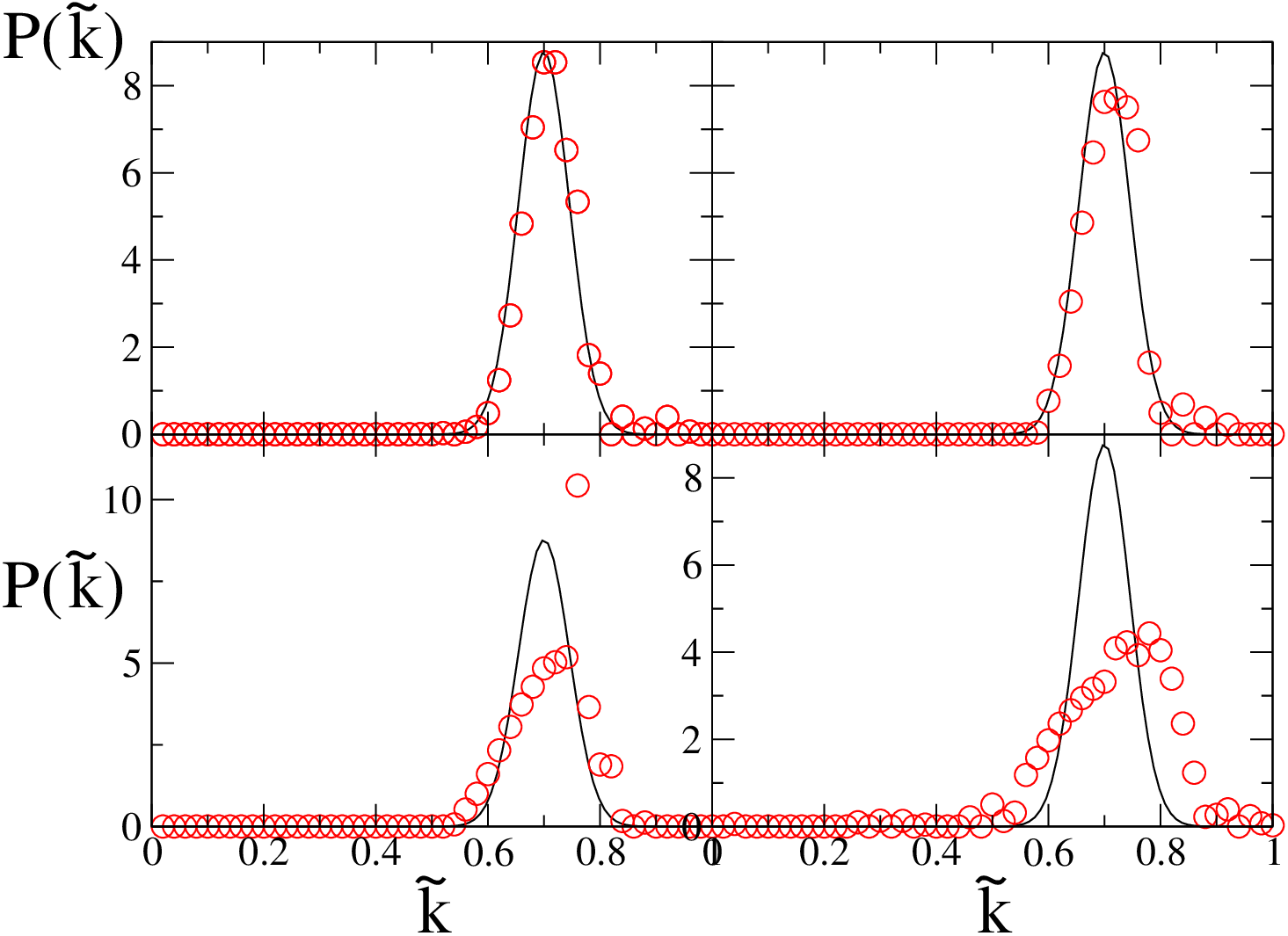}
\caption{
Solution of the inverse problem by the HMF equations in the presence
of noise added to the external currents. We consider the same setup of Fig. 9 
and we compare, for different
values of the noise amplitude $\delta$, the 
reconstructed probability distribution $P(\tilde k)$ (red circles) with the
original gaussian distribution (black line): the upper--left panel 
corresponds to the
noiseless case ($\delta =0$), while the upper--right, the lower--left and 
and the lower--right correspond to $\delta = 0.1, 0.2, 0.3$, respectively.
}
\label{noise_invert_1}
\end{figure}

Also the inversion procedure exhibits the same robustness
with respect to noise. As a crucial test, we have solved the inverse problem 
to recover $P(\tilde k)$ by injecting the noisy signal $Y(t)$  in the noiseless equations 
(\ref{vktil})--(\ref{zktil}), where $a = \bar a$  (see  Fig.\ref{camponoise}). 
The reconstructed distributions $P(\tilde k)$, for different $\delta$, are shown in  Fig. \ref{noise_invert_1}.
For relatively small noise amplitudes ($\delta< 0.1$) the recovered form of $P(\tilde k)$ is quite
close to the original one, as expected because the noisy $Y(t)$ does
not differ significantly from the noiseless one.
 On the contrary, for relatively large noise amplitudes
 ($\delta>0.1$),  the recovered  distribution
 $P(\tilde k)$  is broader than the  original one and centered around
 a  shifted average value $\langle \tilde k \rangle$.
The dynamics exhibits much weaker synchrony effects, the same 
indeed one could observe for the noiseless dynamics on the
lattice built up  with this broader  $P(\tilde k)$ given by the inversion method.

\begin{figure}[htbp]
\centering
\includegraphics[scale=0.35]{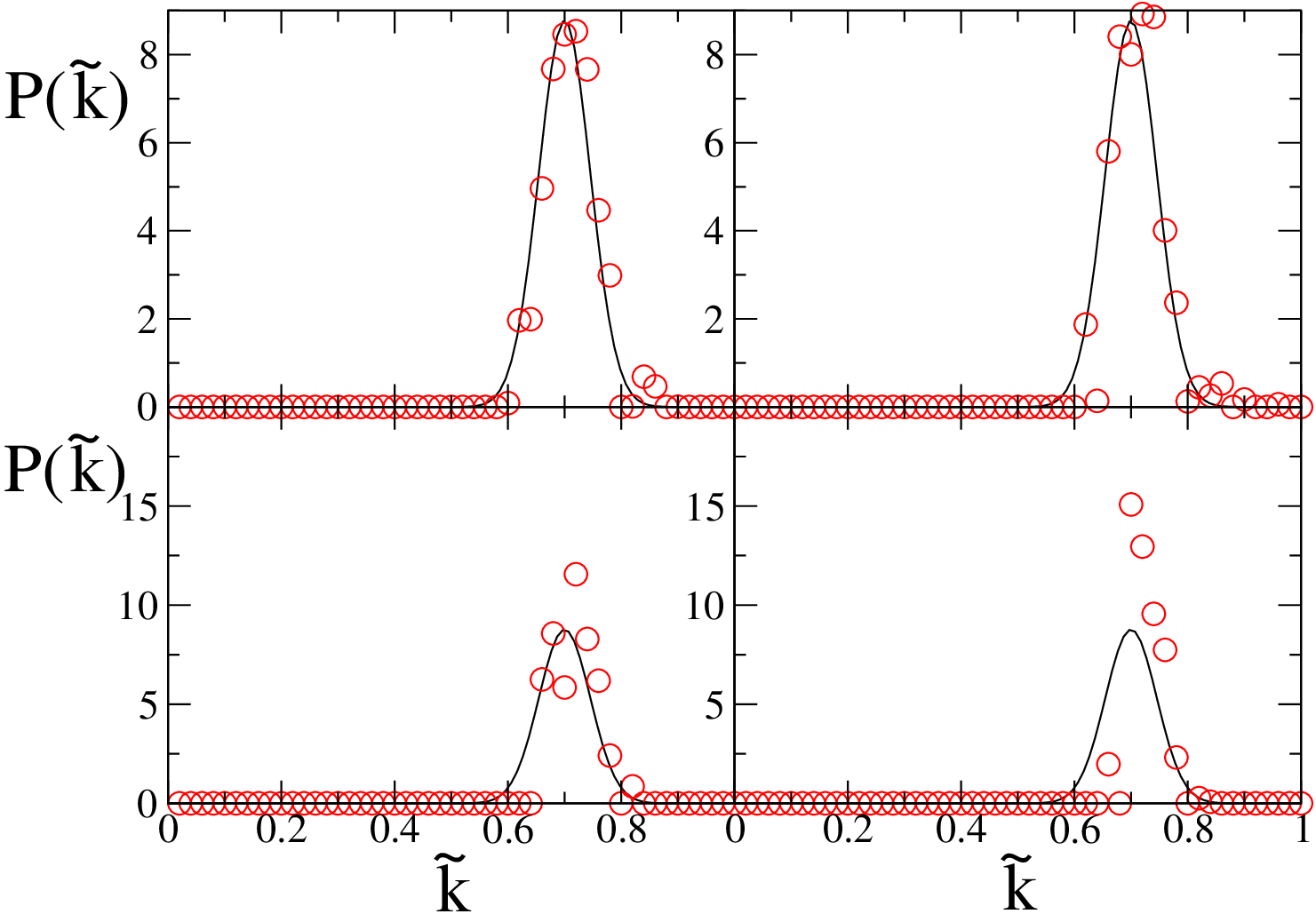}
\caption{
Solution of the inverse problem by the HMF equations in the presence
of noise added to the activity field. We consider the same setup of 
Fig. 9, where now $a = 1$ and $Y_{\delta}(t) = (1 - \eta(t))Y(t)$ (the random
variable $\eta(t)$ is extracted from a uniform probability distribution 
in the interval $[-\delta/2, \delta/2]$).
We compare, for different
values of the noise amplitude $\delta$, the 
reconstructed probability distribution $P(\tilde k)$ (red circles) with the
original gaussian distribution (black line): the upper--left, 
the upper--right, the lower--left and 
and the lower--right panels correspond to $\delta = 0.1, 0.4, 0.8, 1.2$, 
respectively.
}
\label{noise_invert_2}
\end{figure}

As a matter of fact, the global neural activity fields obtained by  experimental
measurements are unavoidably affected by some level of noise. 
Accordingly, it is  worth investigating the robustness of the inversion
method also in the case of noise acting directly on $Y(t)$. 
In order to tackle this problem, we have considered a simple
noisy version of the global synaptic activity field,  defined as
$Y_{\delta} (t) =(1+\eta(t))Y(t) $, where the random number $\eta(t)$  is
uniformly extracted, at each integration time step,  in the interval $[-\frac{\delta}{2},\frac{\delta}{2}]$.
In Fig. \ref{noise_invert_2} we show 
the distributions $P(\tilde k)$ obtained for different values of $\delta$. 
We can conclude  that the inversion method is quite stable with respect to this  
additive noise. In fact, even for very large signal--to--noise ratio (e.g. low--right panel of Fig. \ref{noise_invert_2},
where $\delta = 0.8$) the main features of the original distribution are still recovered,
within a reasonable approximation.

\section{HMF in sparse networks}
\label{sec7}

In this section we analyze the effectiveness of the HMF
approach for sparse networks, i.e. networks where the neurons
degree does not scale linearly with $N$ and, in particular, the average degree
$\langle k \rangle$ is independent of the system size. 
In this context, the coupling term describing
the membrane potential of a generic neuron $i$, in a network of $N$ neurons, 
evolves according to the following equation:
\begin{equation}
\label{vsparsa}
\dot v_j= a -v_j + \frac{g}{\langle k\rangle} \sum_{i \ne j} \epsilon_{ij} y_i, \, 
\end{equation} 
while the dynamics of $y_i$ is the same of Eq.s
(\ref{dynsyn})--(\ref{contz}). The coupling therm is now independent of
$N$, and the normalization factor, 
$\langle k\rangle$, has been introduced in order to compare models with different average connectivity. 
The  structure of the adjacency matrix $\epsilon_{ij}$ is determined 
by choosing for each neuron $i$ its  in-degree $k_i$  from a probability distribution $P(k_i)$
(with support over positive integers)      independent of the system size.

\begin{figure}[htbp]
\centering\
\includegraphics[scale=0.35]{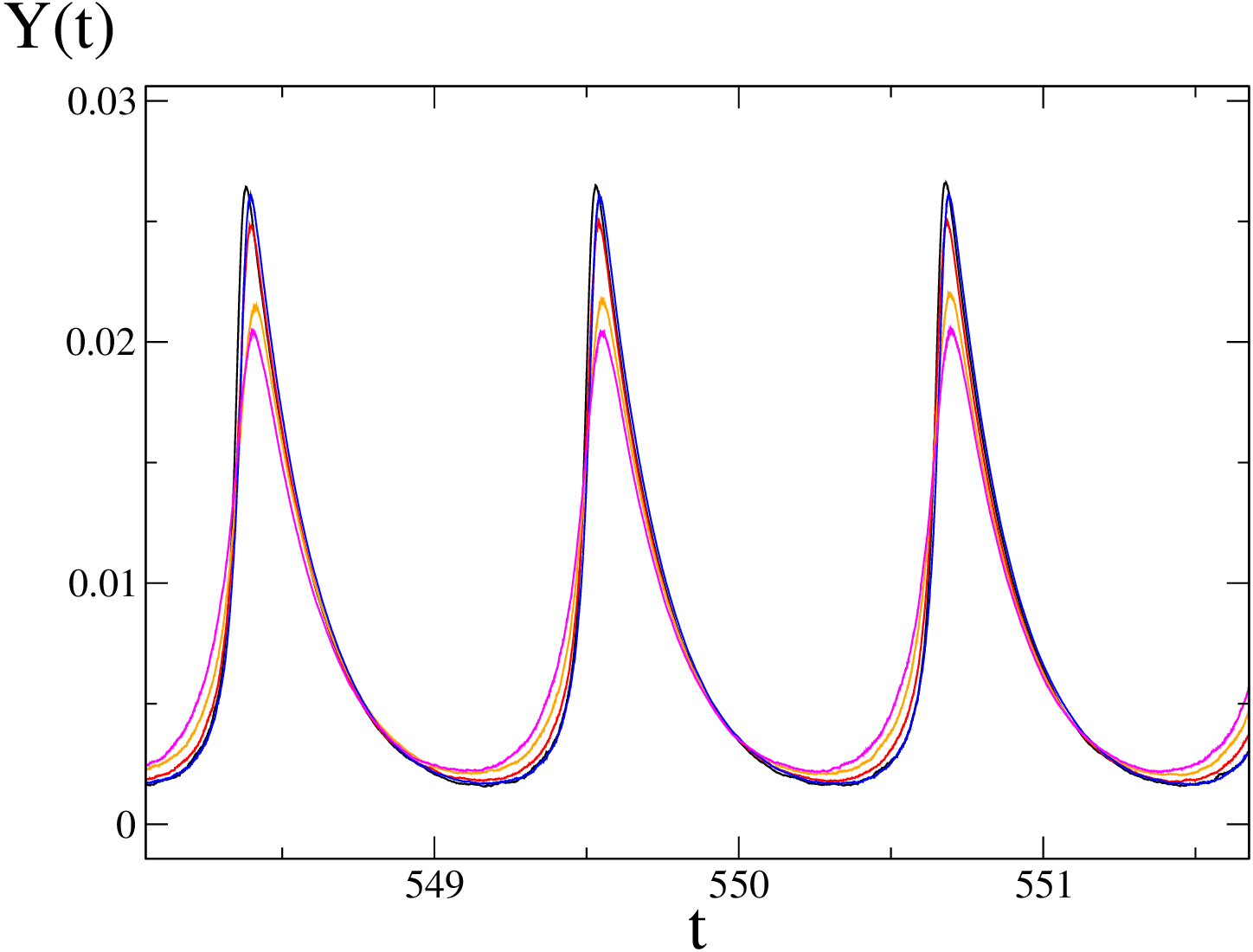}
\vskip 30pt
\includegraphics[scale=0.35]{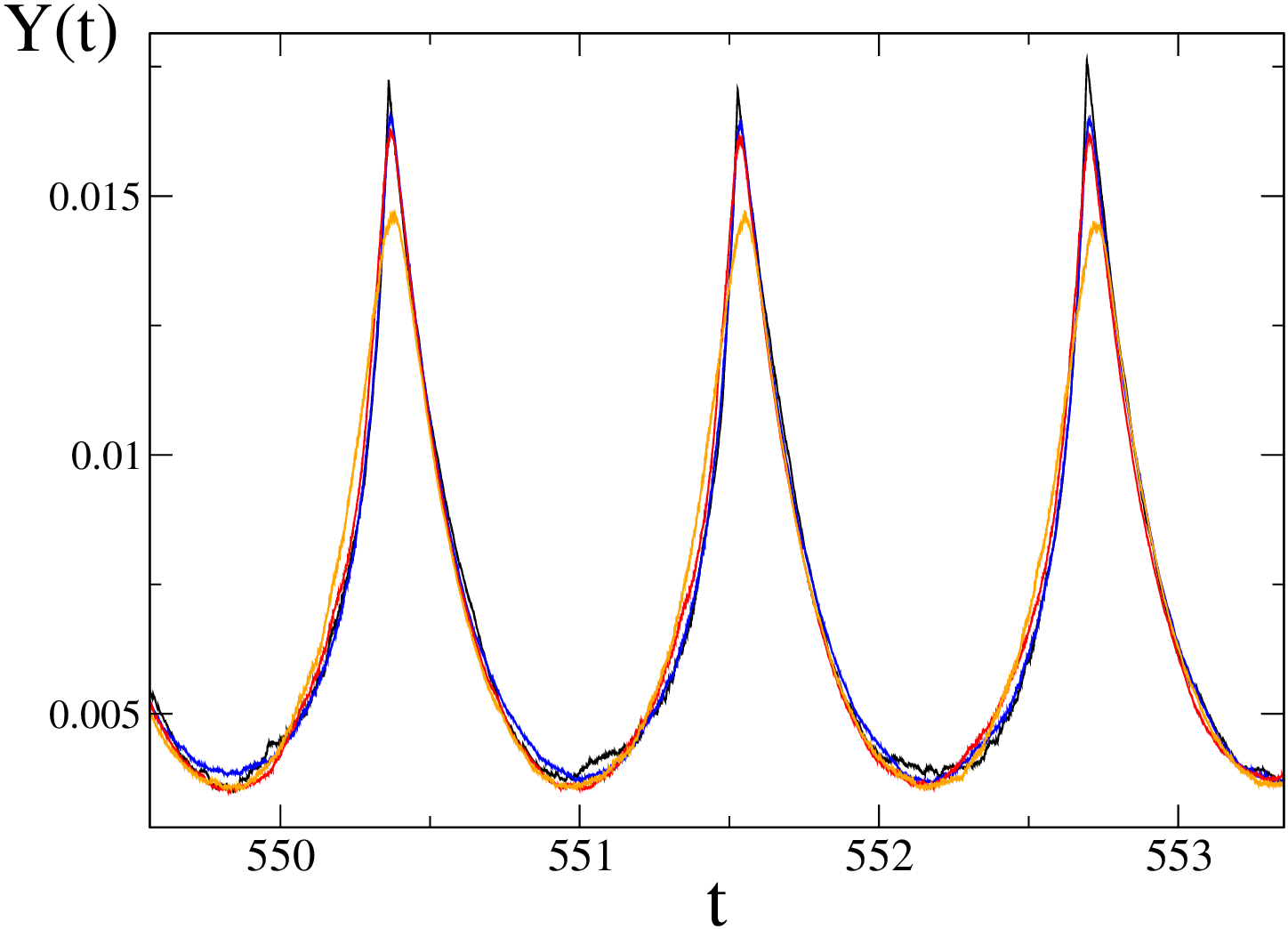}
\caption{
Comparison of the global synaptic activity field $Y(t)$ from 
sparse random networks with the same quantity generated by the corresponding
HMF dynamics. We have considered sparse random networks with 
$N= 10^4$ neurons. In the upper panel  we consider  a
Gaussian probability distributions $P(k)$ with different averages
$\langle k \rangle$ and 
variances $\sigma_k$, such that $ \sigma_k/\langle k \rangle = 0.06$:
$\langle k \rangle = 10, 20, 60, 100$ correspond to the violet, orange, red and blue
lines, respectively. The black line represents $Y(t)$ from the HMF dynamics
($M=10^3$), where $\hat P(\hat k)$ is a Gaussian probability distribution
with $\langle \hat k \rangle = 1$ and $\sigma_{\hat k} = \sigma_k/\langle k 
\rangle = 0.06$.  In the lower panel  we consider  
the scale free case with fixed power exponent $\alpha$ and different $k_m$:
$k_m = 10, 30, 70$ correspond to the orange, red and blue
lines, respectively. The black line represents $Y(t)$ from the HMF dynamics
($M=10^3$), where $\hat P(\hat k)=(\alpha-1)\hat{k}^{-\alpha} $
with cutoff $\hat k_m = 1$.}
\label{campi_sparsa}
\end{figure}

On sparse networks the HMF model is not recovered in the thermodynamic limit, 
as the fluctuations of the field received by each neuron of in--degree $k_i$ 
do not  vanish for  $N\to \infty$.
Nevertheless, for large enough values of $k_i$, one can expect that the
fluctuations become negligible in such a limit,
i.e. the synaptic activity field received by different neurons with the same 
in-degree is approximately the same. 
Eq. (\ref{vsparsa})  can be turned into a mean--field like
form as follows
\begin{equation}
\label{vsparsa_mean}
\dot v_j= a -v_j + \frac{g}{\langle k\rangle} k_jY ~,  
\end{equation} 
where $Y(t)$ represents the global field, averaged over all neurons in the
network. This implies that the equation  is the same for all neurons with 
in--degree $k_j$, depending only on the ratio $\hat{k}_j=k_j/\langle
k\rangle$.  
Consequently, also in this case one can read Eq. (\ref{vsparsa_mean}) as a HMF formulation of Eq. (\ref{vsparsa}), 
where each class of  neurons $\hat{k}$ evolves according to
to Eq.s (\ref{vk})--(\ref{zk}), with $\hat{k}$ replacing $\tilde k$, while the global activity field is given by
the relation 
$Y(t)=\int_0^{\infty}\hat{P}(\hat{k})y_{\hat{k}}(t)d\hat{k}$.  

In order to analyze the validity of the HMF as an approximation of models defined 
on sparse networks, we consider two main cases: ({\sl i})   $\hat{P}(\hat{k})$ is a truncated Gaussian
with average $ \langle \hat{k}\rangle=1$ 
and standard deviation $\sigma_{\hat{k}}$; ({\sl ii}) $\hat{P}(\hat{k})=(\alpha-1)\hat{k}^{-\alpha}$ is a 
power--law (i.e., scale free) distribution with a lower cutoff $\hat{k}_m=1$.
The Gaussian case ({\sl i}) is  an approximation of
any sparse model, where $P(k_j)$ is a discretized Gaussian distribution 
with parameters  $\langle k\rangle$ and $\sigma_k$, chosen in such a way that
$\sigma_{\hat{k}} = \sigma_k/\langle k\rangle$. The scale free case ({\sl ii}) approximates
any sparse model, where $P(k_j)$ is a power law with exponent $\alpha$ and a generic cutoff.  Such an
approximation is expected to provide better results the larger is $\langle k \rangle$, i.e.  
the larger is the cutoff $k_m$ of the scale free distribution.
In Fig. \ref{campi_sparsa} we plot the global field emerging from 
the HMF model, superposing those coming from a large finite size realization 
of the sparse network, with different values of $\langle k\rangle$ for the Gaussian case (upper panel) and of $k_m$ for the scale free case (lower panel).
The HMF equations exhibit a remarkable agreement with models on sparse 
network,  even for relatively small values of $ \langle k\rangle$ and $k_m$. 
This analysis indicates  that the HMF approach works also for
non--massive topologies, provided the typical connectivities
in the network are large enough, e.g. $\langle k\rangle \sim {\mathcal O} (10^2)$
in a Gaussian random network with $N=10^4$ neurons (see Fig. (\ref{campi_sparsa})).

\section{Conclusions}
\label{sec8}

For systems with a very large number of components, the effectiveness
of a statistical approach, paying the price of some necessary approximation, has been
extensively proven, and  mean--field methods are typical in this sense. 
In this paper we discuss how such a method, in the specific form of
Heterogeneous Mean--Field, can be defined in order to fit an 
effective description of neural dynamics on random networks. 

The relative simplicity of the model studied here, 
excitatory  leaky--integrate--and fire neurons with short term
synaptic plasticity, is also a way of providing a pedagogical
description of the HMF  and of its potential interest in similar contexts \cite{BCDLV}.  

We have reported a detailed study of the HMF approach including
investigations on 
{\sl (i)} its stability properties,  
{\sl (ii)} its effectiveness in describing the dynamics and in solving
the associated inverse problem for different network topologies,
 {\sl (iii)} its robustness with respect to noise, 
and {\sl (iv)} its adaptability to different formulations of the model
at hand. In the light of {\sl (ii)}  and  {\sl (iii)}, the HMF approach
appears quite a promising tool to match
experimental situations, such as the identification of topological  features
of real neural structures, through the inverse analysis of signals
extracted as  time series  from small, but not microscopic, domains.
On a mathematical  ground, the HMF approach is a simple and effective
mean--field formulation, that can be extended to other neural network models
and also to a wider class of dynamical models on random graphs.
The first step in this direction could be the extension of  the HMF method
to the more interesting case, where the random network contains 
excitatory and inhibitory neurons, according to distributions of interest
for neurophysiology \cite{abeles, bonif} . This will be the subject of our future work.

\begin{acknowledgments}
R.L. acknowledges useful discussions
with A. Pikovsky and L. Bunimovich.

\end{acknowledgments}

\end{document}